%% file: main.tex
\documentclass[a4paper,12pt]{article}
\usepackage{amsfonts, amsmath, amssymb, amsthm}
\usepackage[toc,page]{appendix}
\usepackage[blocks]{authblk}
\usepackage{array}
\usepackage[english]{babel}
\usepackage{bm}
\usepackage{booktabs}
\usepackage{chngcntr}
\usepackage{enumerate}
\usepackage{epsfig}
\usepackage{float}
\usepackage{fontenc}
\usepackage{graphicx}
\usepackage[hmargin=2cm,vmargin=3cm]{geometry}
\usepackage[colorlinks,allcolors = blue]{hyperref}
\usepackage{lscape}
\usepackage{mathrsfs}
\usepackage{multicol}
\usepackage{multirow,bigdelim}
\usepackage{natbib}
\usepackage{rotating}
\usepackage{subfigure}
\usepackage{verbatim}
\usepackage[svgnames]{xcolor}

\usepackage{mdframed}
\usepackage{graphicx}

\input{colors.tex}
\input{environments.tex}
\input{tools.tex}
\input{short_bib.tex}

\def\vc#1{{\textcolor{black}{{#1}}}}

\date{}
\begin{document}
\renewcommand\thefootnote{}
\title{\vspace{-3.1cm}Bayesian Mixture Models for Heterogeneous Extremes}
\author[$*$]{\footnotesize Viviana \textsc{Carcaiso}}
\author[$\dagger, \ddagger$]{\footnotesize Miguel \textsc{de Carvalho}}
\author[$\S$]{\footnotesize Ilaria \textsc{Prosdocimi}}
\author[$\S$]{\footnotesize Isadora \textsc{Antoniano-Villalobos}}
\affil[$*$]{Unité Biostatistique et Processus Spatiaux, INRAE, France}
\affil[$\dagger$]{School of Mathematics, University of Edinburgh, UK}
\affil[$\ddagger$]{Department of Mathematics, University of Aveiro, Portugal}
\affil[$\S$]{Department of Environmental Sciences, Informatics and Statistics, Ca'Foscari University of Venice, Italy}

\date{}
\maketitle


\vspace{-1.5cm}
\begin{abstract}\footnotesize 
  The conventional use of the Generalized Extreme Value (GEV) distribution to model block maxima may be inappropriate when
  extremes are actually structured into multiple heterogeneous groups. 
In this work, we propose a novel approach for describing the behavior of extreme values in the presence of such heterogeneity.
Rather than defaulting to the GEV distribution simply because it arises as a theoretical limit, we show that alternative block maxima-based models can also align with the extremal types theorem while providing improved flexibility in practice. Our formulation leads us to a mixture model that has a Bayesian nonparametric interpretation as a Dirichlet process mixture of GEV distributions. The use of an infinite number of components enables the characterization of every possible block behavior, while at the same time capturing similarities between observations based on their extremal behavior. 
By employing a Dirichlet process prior on the mixing measure, we 
can capture the complex structure of the data without the need to pre-specify the number of mixture components. 
The application of the proposed model is illustrated using both simulated and real-world data. \vspace{0.2cm}\\
{\footnotesize \textsc{key words:} Bayesian nonparametrics; Dirichlet process mixture; Extreme events; Extreme value theory; Homogeneity; Heterogeneity.}
\end{abstract}
\let\thefootnote\relax\footnotetext{}
\vspace{-0.4cm}

\section{Introduction}\label{introduction}
Extreme value theory \citep{coles2001, beirlant2004, HandbookExtremes2026} provides a statistical framework for modeling the tails of probability distributions. It offers tools to analyze and quantify the probability of extreme events, with the goal of assessing the risk associated with such occurrences. In a block maxima analysis, the data are partitioned into non-overlapping blocks, and the maximum value within each block is considered. Under mild conditions, the resulting series of block maxima is approximately independent and identically distributed, following a Generalized Extreme Value (GEV) distribution as a consequence of the extremal types theorem \citep[][Theorem~3.1]{coles2001}. The GEV distribution plays a central role in extreme value analysis due to this theoretical foundation. However, the theorem relies on the strong assumption that the raw data are independent and identically distributed, an assumption that may not hold in practice. Even when these conditions are met, the GEV distribution is merely a limiting object, whereas in practice only a finite amount of data is available.

In this paper, we propose a flexible alternative model that remains grounded in the extremal types theorem while addressing some potential misspecifications. Our approach generalizes the GEV by allowing the data to reveal when cross-block heterogeneity is present or when the GEV assumption is overly restrictive. We challenge the common practice of defaulting to the GEV distribution solely because it arises as a theoretical limit. As we demonstrate, alternative block-maxima-based models can be constructed that remain consistent with the extremal types theorem while offering a lower degree of misspecification and greater adaptability to diverse data patterns. The reasons underlying the GEV misspecification that our approach mitigates are twofold. First, the GEV arises as a limiting model, whereas our infinite mixture is sufficiently flexible to contain it as a special case; thus, it can accommodate deviations from the asymptotic regime at sub-asymptotic levels. Second, we emphasize that under heterogeneity, each block maximum should be viewed as arising from a distinct GEV component associated with a mixture model, thereby motivating our approach over a single GEV specification.

Our model specification is based on a mixture of GEV kernels and thus contributes to the growing body of work exploring mixture models in extreme value theory. \cite{boldi2007} introduced a mixture model for the angular density in a multivariate extreme value setting. Bernstein polynomials, which can be viewed as flexible mixtures of beta densities, have also been used to model multivariate extremes \citep[e.g.,][]{guillotte2016, marcon2016, hanson2017}. More recently, \cite{tendijck2023} proposed a mixture-based extension of the Heffernan–Tawn model for bivariate extremes, and \cite{andre2025} introduced Gaussian mixture copulas to model dependence in both the body and the tails of joint distributions. Our proposed methodology differs from these contributions in several fundamental respects. Most notably, we focus on the block maxima setting; the kernel in our mixture corresponds to the canonical univariate extreme value distribution, widely regarded as the natural reference in this context; finally, our emphasis is on marginal tail behavior, rather than on extremal dependence. Parenthetically, it is important to distinguish our contribution from models based on spliced distributions, which are sometimes also referred to as extreme value mixture models \citep[e.g.,][]{macdonald2011}---though the term extreme value spliced distributions would arguably be more accurate for the latter. 

Whereas inference via EM algorithms for mixtures of extreme-value distributions has been studied by \cite{otiniano2017}, our work takes a step further by developing the probabilistic and statistical foundations for such mixtures from an extreme value theory perspective. In particular, we formally connect the proposed framework to the extremal types theorem, emphasize its suitability for modeling heterogeneous extremes, and show that it admits a natural Bayesian nonparametric formulation. As a consequence, our contribution also adds to the growing literature on bimodal models for extremes \citep{otiniano2023} as well as to the emerging 
interface between extreme value theory and Bayesian nonparametrics \citep[e.g.,][]{guillotte2011, lugrin2016, hazra2021, decarvalho2024, ramirez2024, leyva2026}. From a Bayesian nonparametric viewpoint, the model may at first appear as a standard Dirichlet process mixture. However, the setting is non-standard due to several features specific to extreme value theory. These include the nonstandard asymptotic regime (block maxima arise as limits in the block size), the parameter-dependent support of the GEV distribution, and the lack of orthogonality in its standard parametrization, which typically complicates inference \citep[e.g.,][]{huet2026}. In line with other block maxima approaches in EVT, the unit of analysis is not the raw data but rather the (heterogeneous) block maximum (e.g., seasonal maxima). Finally, we emphasize that, unlike most of the Bayesian nonparametric literature where the choice of kernel is largely arbitrary as under mild conditions it does not affect the support of the induced mixture model \citep[Section~3.1]{lijoi2004extending}; here, however, the kernel is directly motivated by EVT via the extremal types theorem \citep[Chapter~3]{HandbookExtremes2026} and a block maxima argument, thereby aligning with standard principles of EVT.

The rest of the paper unfolds as follows. Section~\ref{methods} introduces the proposed framework and model. The main findings from a simulation study are presented in Section~\ref{simulations}. Section~\ref{applications} illustrates the methodology with two real data examples. Section~\ref{sec:Final} provides concluding remarks and discusses open problems. Technical details and proofs are included in the Appendix. 

\section{Bayesian Modeling of Heterogeneous Extremes}\label{methods}

\subsection{Heterogeneous Extremes} \label{sec:2.1}
\subsubsection*{Probabilistic Setup}
Let $\mathbb{M}_n = \max\{X_1, \dots, X_n\}$ denote the sample maximum of a random sample $X_1, \dots, X_n \sim F$. The extremal types theorem can be viewed as a counterpart to the central limit theorem for maxima \citep[Chapter~3]{HandbookExtremes2026}; it states that, as $n \to \infty$, if there exist sequences $a_n > 0$ and $b_n \in \mathbb{R}$ such that $(\mathbb{M}_n - b_n)/a_n$ converges in distribution to a nondegenerate limit, then the limit must be an extreme value distribution. That is, 
\begin{equation*}
\lim_{n \to \infty} \Pr\left( \frac{\mathbb{M}_n - b_n}{a_n} \le z \right) = G_\xi(z), \quad z \in \mathbb{R}. 
\end{equation*}
We say that $F$ belongs to the maximum domain of attraction of an extreme value distribution $G_\xi$, denoted $F \in D(G_\xi)$; its distribution function is given by
\begin{equation}\label{eq:G}
  G_{\xi}(z) = \exp\{ - (1 + \xi z)^{-1/\xi}\},
  \quad 1 + \xi z > 0,
\end{equation}
and $G_{\xi}(z) = 0$ otherwise, with $G_{\xi}(z) \to 1$ as $z \to z_+$, where $z_+ = \infty$ if $\xi \ge 0$ and $z_+ = -1/\xi$ if $\xi < 0$. Here, $\xi \in \mathbb{R}$ is the shape parameter (extreme value index) governing the tail behaviour; the three classical types correspond to $\xi > 0$ (Fr\'echet), $\xi < 0$ (Weibull), and $\xi = 0$ (Gumbel).

In practice, the normalizing constants $a_n$ and $b_n$ are learned from data, resulting in the inclusion of location and scale parameters in \eqref{eq:G}. This motivates the GEV distribution whose distribution function is 
\begin{equation}\label{eq:GEV}
  G(z \mid \mu, \sigma, \xi) =
  \exp \left\{ - \left[ 1+ \xi \left( \frac{z-\mu}{\sigma} \right) \right]^{-1/\xi} \right\},
\end{equation}
for $z \in \mathcal{Z} = \{ z \in \mathbb{R} : 1 + \xi (z-\mu)/\sigma > 0 \}$, and $G(z \mid \mu, \sigma, \xi)=0$ otherwise, where $\mu \in \mathbb{R}$ and $\sigma > 0$ are the location and scale parameters, respectively; moreover, $G(z \mid \mu, \sigma, \xi) \to 1$ as $z \to z_+$, where $z_+ = \infty$ if $\xi \ge 0$ and $z_+ = \mu - \sigma/\xi$ if $\xi < 0$.

The setting above corresponds to the classical setup, which we refer to as homogeneous extremes. We now introduce the setting of interest, which we refer to as heterogeneous extremes. Assume that a latent indicator $C$ supported on $\{1, \dots, K\}$ identifies the group to which a sample belongs. Formally,
\begin{equation}\label{setup}
  (X_{k,1}, \dots, X_{k,n} \mid C = k)~\iid~F_k \in D(G_{\xi_{k}})
\end{equation}
is a random sample of size $n$ from group $k$; for simplicity, we assume that each sample has the same length but the argument below applies more generally. Define the heterogeneous sample maximum as the maximum indexed by the value of $C$ corresponding to its originating sample, that is,
\begin{equation}\label{hsm}
  M_n = \sum_{k=1}^K I(C=k)\, M_{k,n} =
  \begin{cases}
    M_{1,n}, & \text{if } C = 1, \\
    ~~~\vdots \\ 
    M_{K,n}, & \text{if } C = K,
  \end{cases}\vs \vs 
\end{equation}
where $I(\cdot)$ is the indicator function and $M_{k,n}= \max\{X_{k,1}, \dots, X_{k,n}\}$ is the maximum of the sample belonging to group $k$. To allow for different domains of attraction in different groups, we define the normalizing sequences of random variables $A_n >0$ and $B_n$ to be group-dependent, that is,
\begin{gather}\label{gdepcons}
   A_n = \sum_{k=1}^K I(C=k) \, a_{k,n}, \qquad B_n = \sum_{k=1}^K I(C=k) \,b_{k,n},
\end{gather}
where $a_{k,n} > 0$ and $b_{k,n}$ are the normalizing sequences for $M_{k,n}$. 
Then, by the law of total probability and the extremal types theorem 
\begin{align}
    \text{P} \left ( \frac{M_n-B_n}{A_n}  \le z \right) &= \sum_{k=1}^K \text{P}(C=k) \, \text{P} \left ( \frac{M_n-B_n}{A_n}  \le z  \mid C=k \right )\label{eq:1F}\\
    &= \sum_{k=1}^K \pi_k \, \text{P} \left ( \frac{M_{k,n}-b_{k,n}}{a_{k,n}}  \le z \right) \xrightarrow[n \to \infty]{}  \sum_{k=1}^K \pi_k G_{\xi_k}(z), \quad z \in \mathbb{R}, \label{eq:F}
\end{align}
where $\pi_k = \Pr(C = k)$. In simpler terms, the properly standarized heterogeneous sample maximum converges in distribution to a mixture of EV distributions as $n \to \infty$. This motivates the following extreme value mixture model for heterogeneous extremes, 
\begin{equation}\label{ourmodel}
  F(z) = \sum_{k=1}^K \pi_k \, G(z \mid \mu_k, \sigma_k, \xi_k), \quad z \in \mathbb{R},
\end{equation}
where $(\mu_k,\sigma_k,\xi_k)^{\T}$ are component-specific parameters that induce heterogeneity across mixture components and $G(z \mid \mu_k, \sigma_k, \xi_k)$ is their corresponding GEV distribution function.


Unlike the standard arguments leading to the GEV distribution \citep[e.g.][Chapter~3]{coles2001, HandbookExtremes2026}, the model proposed here is derived from the heterogeneous sample maximum $M_n$ which, in contrast to the classical sample maximum, incorporates an additional layer of randomness induced by $C$. This extra source of randomness, absent in the sample maximum $\mathbb{M}_n$, is the key mechanism driving the mixture in \eqref{eq:F}.

Some final remarks on technical details are in order. Firstly, when $K$ is random, the limit in \eqref{eq:F} has to be interpreted in the sense of almost sure convergence; also, as shown in the Appendix, the result in \eqref{eq:F} is also valid when $K \to \infty$ as a consequence of the well-known Tannery's lemma \citep{tannery1910introduction}. 


\subsubsection*{Why Cross-Block Heterogeneity Is Better Represented by our Mixture Model}

The standard rationale for block maxima focuses on the need for $n$ to be sufficiently large for the GEV approximation to hold. However, an equally important question concerns \emph{which} GEV distribution is being approximated, since under cross-block heterogeneity each block may correspond to a different GEV distribution. To see this, suppose we observe $m := m_n$ homogeneous blocks 
\begin{equation}\label{blocks}
  \{B_1 ,\dots, B_m\}, 
\end{equation}
where $B_i = \{X_{c_i,1}, \dots, X_{c_i,n}\}$ and $c_1,\dots,c_m \iid F_C$ with $F_C$ denoting the cumulative mass function of $C$; again, for simplicity \eqref{blocks} assumes equal block lengths, but the arguments below apply more generally provided each block is sufficiently large.

For sufficiently large $n$, the extremal types theorem implies that the $i$th block maximum is approximately GEV-distributed, that is,
\begin{equation}\label{blockm}
  Z_i := M_{c_i} = \max\{B_i\}~\dot\sim~\mathrm{GEV}(\mu_{c_i},\sigma_{c_i},\xi_{c_i}), 
  \qquad i = 1,\dots,m,
\end{equation}
as $n \to \infty$. It follows from \eqref{blockm} that under cross-block heterogeneity it is inappropriate to treat the block maxima $Z_1,\dots,Z_m$ as identically distributed. In particular, fitting a single GEV to the data is not justified; instead, each $Z_i$ should be viewed as the block maximum associated with the component $C=c_i$. The following example illustrates the framework. 

\begin{figure}
  \centering
  \includegraphics[scale = 0.5]{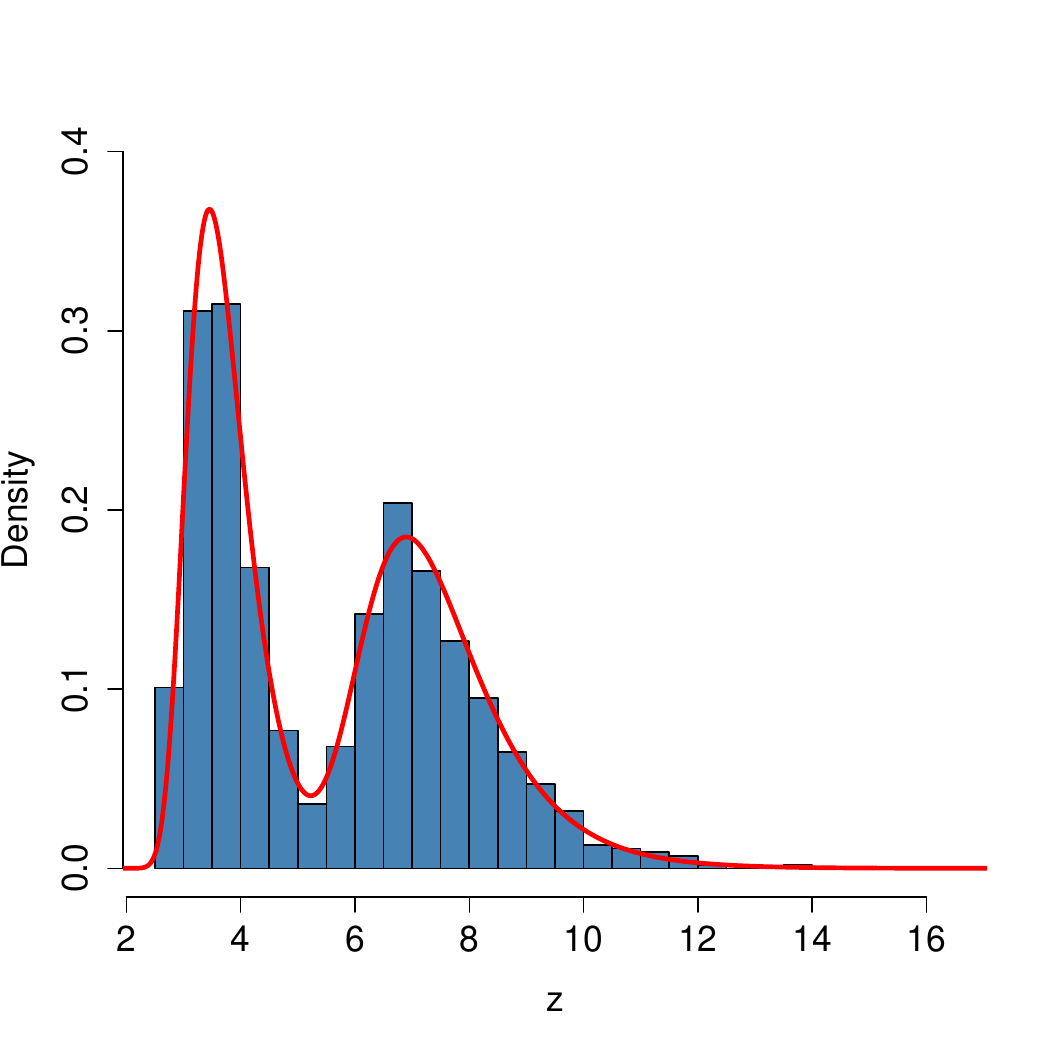}
  \caption{\footnotesize \label{example} Density of two-component block maxima based on Example~\ref{2compex}. Fifty blocks are generated from two exponential distributions, one with $\lambda = 1$ and the other with $\lambda = 2$; each block has size $n = 1000$. The red line represents the density that follows from \eqref{eq:F}.}
\end{figure}

\begin{example}[Two-component block maxima]\label{2compex} \normalfont 
We illustrate cross-block heterogeneity using block maxima from two exponential distributions.
Let $C \in \{1,2\}$ with $\pi_k = \text{P}(C = k) = 1/2$, for $k \in \{1,2\}$, and let $F_k(x) = 1 - \exp(-\lambda_k x)$ be the distribution function of an exponential distribution with rate parameter $\lambda_k > 0$, for $k = 1,2$. Fig.~\ref{example} illustrates an example where $m = 50$ blocks are generated from this model with $\lambda_1 = 1$ and $\lambda_2 = 2$; each block has size $n = 1000$. The histograms of these heterogeneous block maxima are compared to the density of 
\begin{equation}\label{dsen}
    F(z) = \frac{1}{2} G(z \mid \log n, 1, 0) + \frac{1}{2} G(z \mid \tfrac{1}{2} \log n, \tfrac{1}{2}, 0),
\end{equation}
which follows by taking account of \eqref{eq:F}  and the standardizing constants for the exponential model \citep[e.g.,][Chapter~3]{embrechts2013}. As seen in Fig.~\ref{example}, there is clear cross-block heterogeneity in the block maxima, which is captured by a two-component mixture, as predicted by \eqref{eq:F}. \strut \hfill$\blacktriangle$
\end{example} 
\subsubsection*{Other Key Insights}
In practice, the structure in \eqref{blocks} is rarely observed exactly. However, as we illustrate below, it is less restrictive than it may appear, and the proposed model remains applicable more broadly, provided each block is sufficiently large for the GEV approximation to hold. To fix ideas, interpret each block in \eqref{blockm} as a season (e.g., in temperature data), with the randomness in $C$ capturing, for example, variation across summers in different years, acting as a season-level random effect. From this perspective, the heterogeneous block maxima in \eqref{blockm} correspond to seasonal maxima. Aggregating at the yearly level (i.e., considering annual rather than seasonal maxima) yields
\begin{equation*}
\{B_1^*, \dots, B_{m/4}^*\},
\end{equation*}
where $B_i^* = \{X_{c_i,1}, \dots, X_{c_i,n}, \dots, X_{c_{i+3},1}, \dots, X_{c_{i+3},n}\}$, assuming for simplicity that $m$ is a multiple of $4$. Each aggregated block $B_i^*$ is then heterogeneous. Nevertheless, for sufficiently large $n$, the corresponding within-block maxima remain approximately GEV-distributed, that is, 
\begin{equation}\label{blockm2}
Z_i^* := \max\{B_i^*\} \,\dot{\sim}\, \mathrm{GEV}(\mu_{c_i}^*,\sigma_{c_i}^*,\xi_{c_i}^*), \quad i = 1, \dots, m/4, 
\end{equation}
but the parameters now correspond to an aggregated regime $C^* = c_i$ that takes values on ${1, \dots, K^*}$, with $K^* \leq K$. As a result, the effective number of components in \eqref{ourmodel} depends on the block size: larger blocks merge distinct regimes and reduce the apparent number of components by averaging their tail behaviour. Once again, it follows from \eqref{blockm2} that under cross-block heterogeneity it is questionable to treat the block maxima $Z_1^*,\dots,Z_m^*$ as identically distributed. In particular, fitting a single GEV model may be more appropriate in the setting of \eqref{blockm2} than \eqref{blockm}, but remains unjustified; instead, each $Z_i^*$ should be viewed as the block maximum associated with the component $C^* = c_i^*$, thereby motivating our approach over a single GEV specification.

\subsection{Bayesian Modeling of Cross-Block Heterogeneity}\label{sec:2.2}
\subsubsection*{Nonparametric Framework}
Let $Z_1, \dots, Z_m$ be a sequence of heterogeneous block maxima, i.e., maxima over $m$ blocks of $n$ independent and identically distributed random variables, where each block is associated with a possibly different group $k \in \mathbb{N}$. Following the arguments in Section~\ref{sec:2.1}, the block maxima approximately follow the mixture representation in \eqref{ourmodel}; to model this in a Bayesian nonparametric framework, we rewrite it as 
\begin{equation}\label{eq:GEV_DP}
    F(z) =  \int_\Theta G(z\mid \mu, \sigma, \xi) H(\text{d}\mu, \text{d}\sigma, \text{d}\xi),
\end{equation}
where $\Theta = \mathbb{R} \times (0, \infty) \times (-1/2, \infty),$ and $H$ is an almost surely discrete random probability measure. 
A Bayesian treatment of heterogeneous extremes entails setting a prior for the mixing measure $H$. A natural option is  the stick-breaking representation \citep{sethuraman1994constructive, ishwaran2001gibbs}:
\begin{equation}\label{eq:SB}
    H = \sum_{k=1}^\infty \pi_k \delta_{\boldsymbol{\theta}_k}, \quad\boldsymbol{\theta}_k \stackrel{\text{iid}}{\sim} H_0,
\end{equation}
where $\pi_k = V_k \prod_{j<k}(1 - V_j)$, with $V_k \sim \text{Beta}(\alpha_k,\beta_k)$ and $\boldsymbol{\theta}_k=(\mu_k, \sigma_k, \xi_k)^{\T}$ for every integer $k$. For the sake of concreteness, we concentrate on the case with $V_k \sim \text{Beta}(1, \alpha)$, in which case $H$ follows a Dirichlet process \citep{ferguson1973bayesian, ferguson1974prior} with precision parameter $\alpha >0 $ and baseline measure $H_0$. Thus, our proposed model for cross-bock heterogeneity can be understood as a Dirichlet process mixture \citep{escobar1995bayesian} with a GEV kernel. More specifically, the proposed model is
\begin{equation}\label{eq:mod}
    \begin{gathered}
        f(z) = \sum_{k=1}^\infty \pi_k \, g(z \mid \boldsymbol{\theta}_k),\\
        \boldsymbol{\theta}_k \mid H_0 \overset{\text{iid}}{\sim} H_0, \quad \pi_k = V_k \prod_{j<k}(1 - V_j), \\
        V_k \mid \alpha \overset{\text{iid}}{\sim} \text{Beta}(1, \alpha),
    \end{gathered}
\end{equation}
where $g$ is the density of the GEV distribution. An additional layer in the model hierarchy can be introduced by assigning a prior to $\alpha$. A conjugate choice is $\alpha \sim \text{Gamma}(a_{\alpha}, b_{\alpha})$.
The proposed model can be fitted using Markov chain Monte Carlo (MCMC) methods \citep[e.g.,][Ch.~2]{reich2019}; technical details of the algorithm are provided in the supplementary materials.

\subsubsection*{Return Levels and Model Diagnostics}
In extreme value theory a key point involves the estimation of return levels, which represent the 
level expected to be exceeded, on average, once in a specific return period (e.g., in 100 years). These levels essentially correspond to the quantiles of a distribution designed for modeling extremes, providing crucial information for risk assessments and decision-making. 
Using parameter estimates, we can derive extreme quantile estimates for the grouped block maxima distribution $F(z) = \sum_{k=1}^\infty \pi_k g(z \mid \boldsymbol{\theta}_k)$ by inverting it. Since the inverse function $F^{-1}$ has no closed form, quantiles need to be numerically computed by finding the values $z$ that solve $F(z)=p$ for various choices of $p$. Typically the return level $r_p$, where $F(r_p)=1-p$, is plotted against $\log\{-\log(1-p)\}$, for selected $p \in (0,1)$. This plot, known as a return level plot, facilitates the detection of the risk associated with extreme events based on the chosen model and the assessment of model performance.

In addition to return levels, quantile residuals \citep{dunn1996} can be readily defined for our model. The key idea behind these residuals is the fundamental fact that if the model fit (e.g., the posterior median $\hat{F}$) is reasonable, then $\hat{F}(Z_i)$ should follow a standard uniform distribution for all $i \in \{1, \dots, m\}$. Consequently, under a well-fitting model,  
\[
\varepsilon_i \equiv \Phi^{-1}(\hat{F}(Z_i))
\]  
is approximately normally distributed for all $i$, whereas deviations from normality indicate a potential poor fit. Here, $\Phi$ denotes the standard normal distribution function, and we refer to the $\varepsilon_i$ as Dunn--Smyth residuals. Normal QQ plots or QQ boxplots \citep{rodu2022} of the observed residuals provide a natural diagnostic tool for assessing the model fit. 




\section{Simulation Study}\label{simulations}
\subsection{Simulation Setup and Preliminary Findings} \label{sec:3.1}
We evaluate the performance of the proposed methods through a simulation study. Here, we describe the data-generating processes and present findings from single-sample experiments; full Monte Carlo study is presented in Section~\ref{sec:3.1}. \vc{Results of an additional simulation scenario based on Example~\ref{2compex} can be found in the supplementary material.}

\begin{table}[H]\caption{\footnotesize Simulation scenarios. Here, $g$, $\phi$, and $t$ are the density functions of the GEV, Normal, and Student's $t$ distributions, respectively.} \label{tab:scenarios}   \vspace{0.2cm}
\centering 
\begin{tabular}{@{}cccc@{}}
\toprule  
\footnotesize \textbf{Scenario} & \footnotesize \textbf{Density function} & \footnotesize \textbf{Parameters} \\ \midrule
\rule{0pt}{1ex} \footnotesize A                 &  \footnotesize $g(z; \mu, \sigma, \xi)$   &   \begin{tabular}[c]{@{}c@{}} \footnotesize $\mu = 10, \sigma = 1.5, \xi =0.2$      \end{tabular}            \\ \hline
\rule{0pt}{5ex} \footnotesize B                 & \footnotesize $ \pi g(z; \mu_1, \sigma_1, \xi_1) + (1-\pi) g(y; \mu_2, \sigma_2, \xi_2)$ &        
\begin{tabular}[c]{@{}c@{}} \rule{0pt}{2ex} \footnotesize $\pi= 0.7$  \\  \rule{0pt}{3ex} \footnotesize $\mu_1=\phantom{8}1  , \sigma_1 = 1.5, \xi_1 = -0.2$   \\  \rule{0pt}{3ex} \footnotesize $\mu_2 = 18, \sigma_2= 1.0, \xi_2 = \phantom{-}0.4$ \end{tabular} 
\\ \hline
\rule{0pt}{7ex} \footnotesize C                 &     \footnotesize  $ \sum_{j=1}^2\pi_j \phi(z; \mu_j, \sigma_j) +\pi_3t(z;\nu)$          &     \begin{tabular}[c]{@{}c@{}}\rule{0pt}{3ex}\footnotesize $\pi_1 =0.25 , \pi_2 = 0.15, \pi_3 =0.6$\\\rule{0pt}{3ex} \footnotesize $\mu_1=4, \sigma_1 =1$ \\\rule{0pt}{3.5ex} \footnotesize $\mu_2=7, \sigma_2 =1$\\\rule{0pt}{3ex} \footnotesize $\nu = 10$\end{tabular}   
\\ \bottomrule
\end{tabular}
\end{table}

We examine the performance of the proposed GEV mixture model across three simulation scenarios. First, we assess its performance when the data are generated from a single GEV distribution (Scenario~A). Next, we assess the model when data are generated from a mixture of two GEV distributions with negative and positive shape parameters (Scenario~B). We then examine performance when data are generated from a mixture of Normal and Student’s $t$ distributions (Scenario~C), to challenge the model in a setting where the data-generating distributions are not typically used for modeling extremes. Table \ref{tab:scenarios} provides an overview of these three data-generating processes, including the specific parameter values. 

For each scenario a random sample of size $1000$ is simulated from the corresponding data-generating process. The blocked Gibbs sampler with fixed truncation \citep{ishwaran2001gibbs, ishwaran2002} described in the supplementary materials  is applied for 30\,000 iterations, with the first half as burn-in. \vc{The algorithm updates at every iteration the parameters of the GEV distribution for each component, the mixing weights and the allocation indicators associating observations to components. The parameters of the GEV are updated using an adaptive Metropolis--Hastings step}.
\vc{In this simulation study} the number of mixture components is truncated at $K=50$, which is high enough compared to the resulting number of occupied components for our finite sample. Importantly, truncation is not equivalent to fitting a standard finite mixture model, as $K$ serves only as an upper bound on the number of components rather than the actual number of occupied ones; see \citet[][Section~2.2]{dunson2010nonparametric}. 

The prior assigned to the parameter of the GEV distributions $\boldsymbol{\theta}_k=(\mu_k, \sigma_k, \xi_k)^{\T}$ in \eqref{eq:mod} is set to be wide, 
\begin{equation*}
  \mu_k \sim \text{N} (0, 10^4), \quad \sigma_k \sim \text{N} (0, 10^4), \quad \xi_k \sim \text{N} (0, 100),  
\end{equation*} 
for every $k$; the precision parameter $\alpha$ is also updated at every iteration as explained in the supplementary materials  (step c2), based on a Gamma prior with $a_{\alpha}=1$ (shape) and $b_{\alpha}=1$ (rate).


\begin{figure}[]
  \vspace{-1.3cm}
    \centering
    \centering
    \footnotesize \textbf{Scenario A} \vspace{0.2cm} \\ 
    \includegraphics[scale = 0.55]{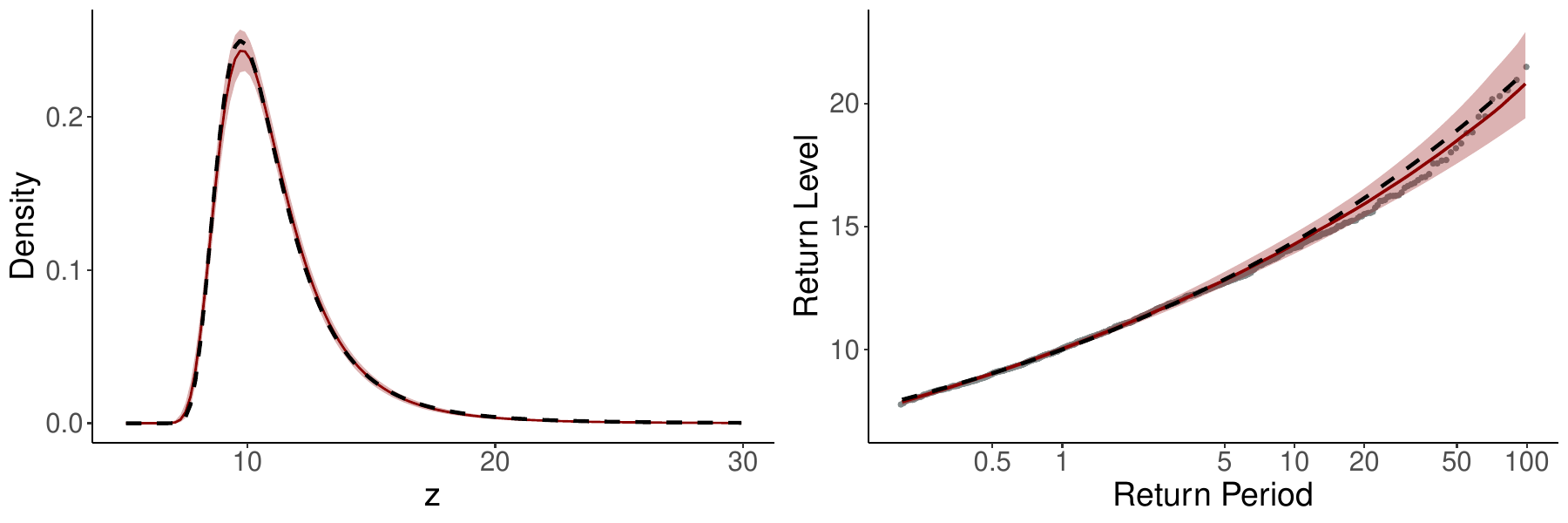}
    \centering
    \footnotesize \textbf{Scenario B} \vspace{0.2cm} \\ 
    \includegraphics[scale = 0.55]{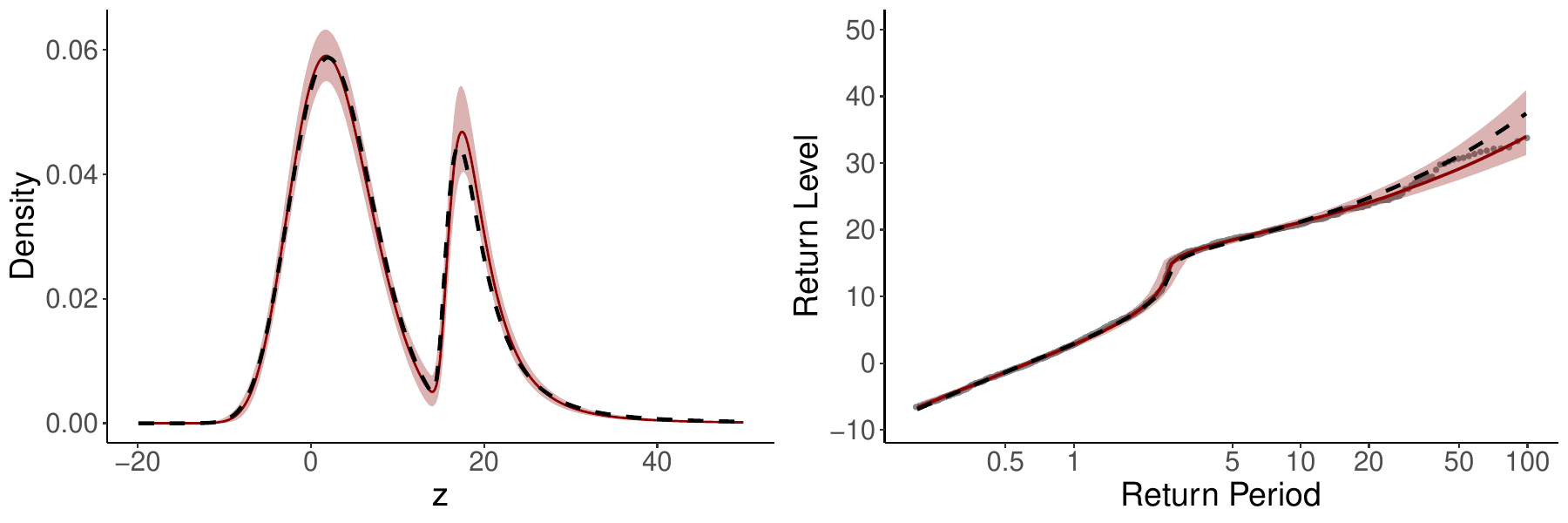}
    \centering
    \footnotesize \textbf{Scenario C} \vspace{0.2cm} \\ 
    \includegraphics[scale = 0.55]{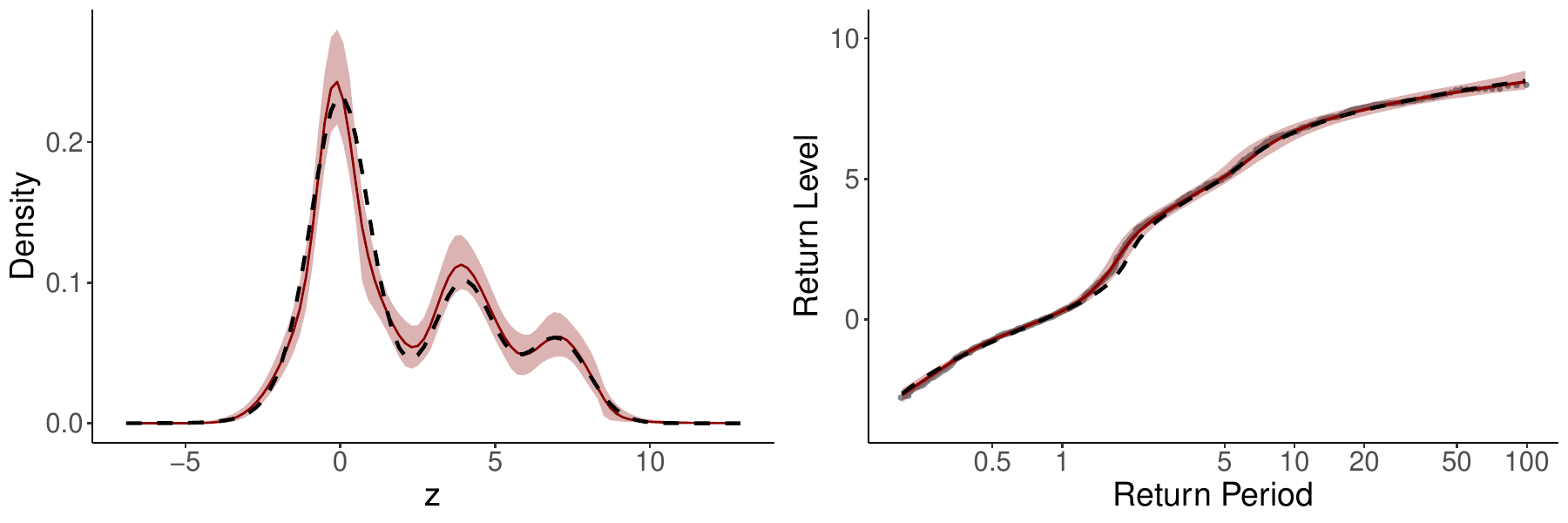}
        \vspace{-0.5cm}
        \caption{\footnotesize Single-sample experiment. Left: posterior median density (solid line) with 95\% credible interval (shaded), compared to the true density (dashed line). Right: posterior median return level curve (solid line) with credible interval (shaded), true return levels (dashed line) and empirical quantiles (gray points).}
        \label{fig:oneshot}
\end{figure}

Fig.~\ref{fig:oneshot} shows results of single-sample experiments for all the scenarios.
On the left, the posterior density from the model is compared to the true one to assess the overall fit. On the right, return level plots are presented to highlight the tail behavior. Posterior median return levels and credible intervals are computed based on a sample of 100 parameter-values, and are compared to the true ones and to the empirical quantiles. 
In Scenarios A and B the model performs very well, \vc{representing both the true} density and quantile levels accurately, as expected. \vc{The slight inaccuracy at the highest return periods seems due to the behavior of the specific sample of data. Uncertainty increases with return period as in both scenarios there are fewer observations in the upper tail.}
In Scenario C the model captures the overall behavior with three modes and the return levels quite well, despite being \vc{in a stronger case of misspecification}.

\vc{While the proposed model performs well in this single-sample experiment, this could be driven by the specific simulated dataset, and we therefore proceed to analyze multiple simulated datasets in the next section.}

\subsection{Monte Carlo Simulations}

\begin{figure}[]
\centering \vspace{-1.3cm}
    \centering
    \footnotesize \textbf{Scenario A} \vspace{0.2cm} \\    
    \includegraphics[scale = 0.55]{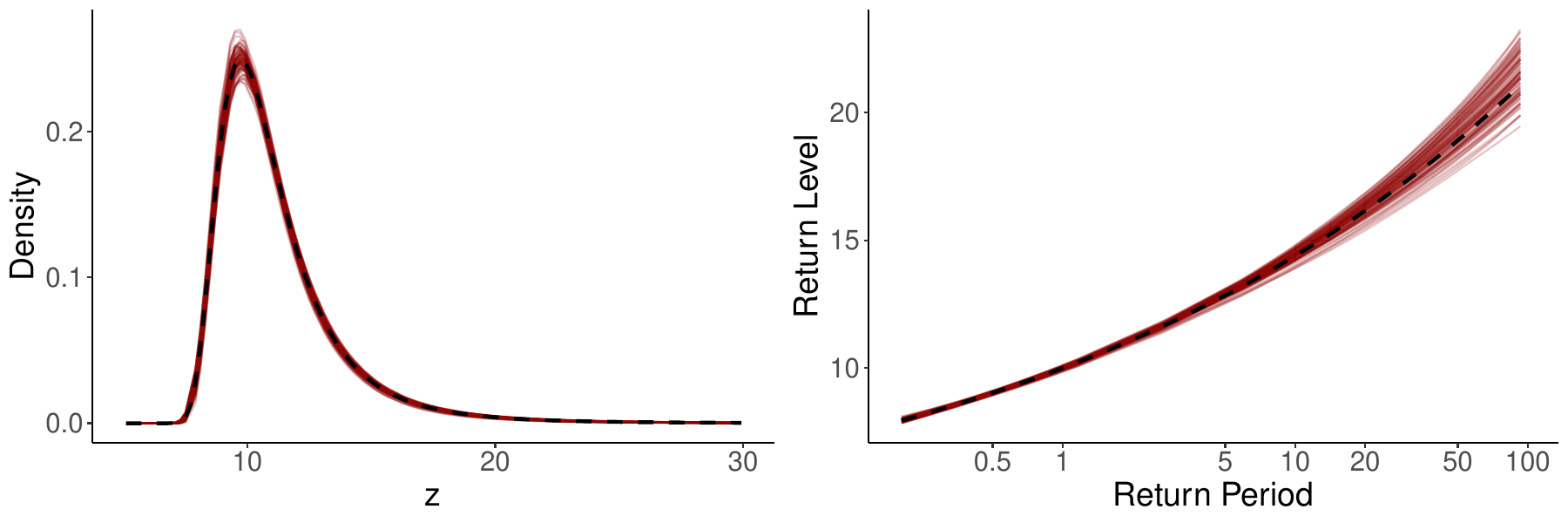}
    \centering
    \footnotesize \textbf{Scenario B} \vspace{0.2cm} \\ 
    \includegraphics[scale = 0.55]{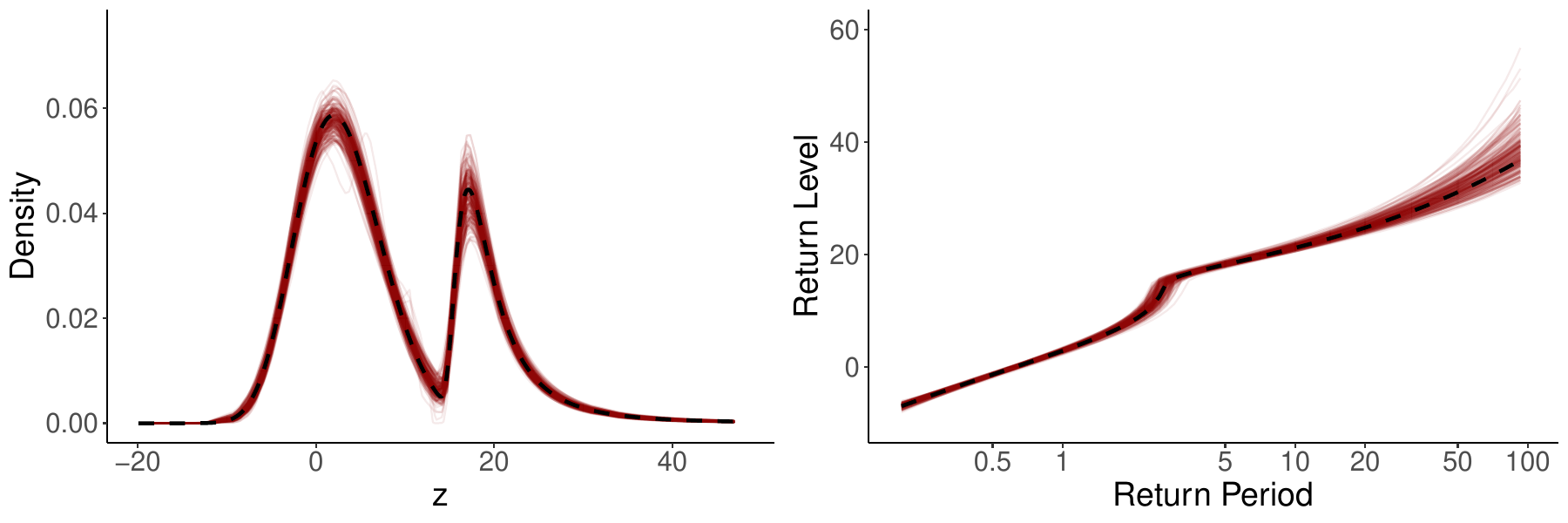}
    \centering
    \footnotesize \textbf{Scenario C} \vspace{0.2cm} \\     
    \includegraphics[scale = 0.55]{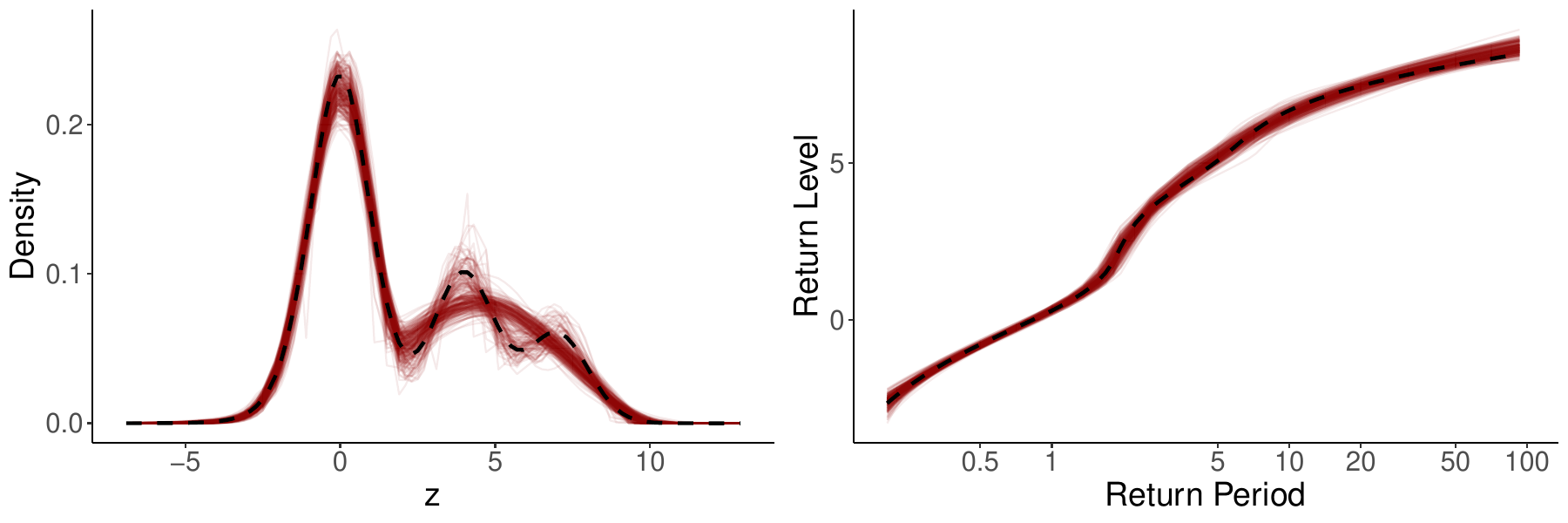}

        \caption{\footnotesize Monte Carlo simulation results, I. Left: median posterior densities. Right: return level curves. The dashed lines represent the corresponding true targets.}
        \label{fig:MC}
\end{figure}

To validate the promising results from the single-sample experiments, a Monte Carlo simulation study is conducted. This involves replicating the one-sample experiments for $M=250$ datasets simulated under each scenario. Fig.~\ref{fig:MC} presents the Monte Carlo median densities and return levels for each scenario, compared with the true values. The findings from the previous one-sample analysis are supported by this Monte Carlo study. \vc{In Scenario A the model recovers very well the true single GEV density and the corresponding return levels for all the samples, with the expected increased variability for high return periods. In Scenario B the density is again well captured;} however the mixture model has a slight tendency to overestimate \vc{the highest return levels}. In Scenario C, which is particularly challenging for the model to fit accurately, \vc{the density plot shows that} there is a noticeable difficulty in separating the \vc{two} components with the smallest data proportions, \vc{which are only recognized in a portion of the simulated samples}, but the return plot shows an overall good fit \vc{for all the samples}. 

To assess the frequentist properties of the Bayesian procedure we use the mean integrated squared error (MISE), given by
    $\text{MISE} = \text{E}[  \int_{\mathbb{R}} \{f(z) - \hat{f}(z)\}^2 \text{d}z],$
where $f$ is the true density under \eqref{eq:mod}, and $\hat f$ is the median posterior density under \eqref{eq:mod}.  

\begin{figure}[H]
  \centering \hspace{1.1cm}
  \footnotesize \textbf{Scenario A} \hspace{3.4cm} 
  \textbf{Scenario B} \hspace{3.4cm} 
  \textbf{Scenario C} \hspace{3.4cm} \\  
  \includegraphics[width=\linewidth]{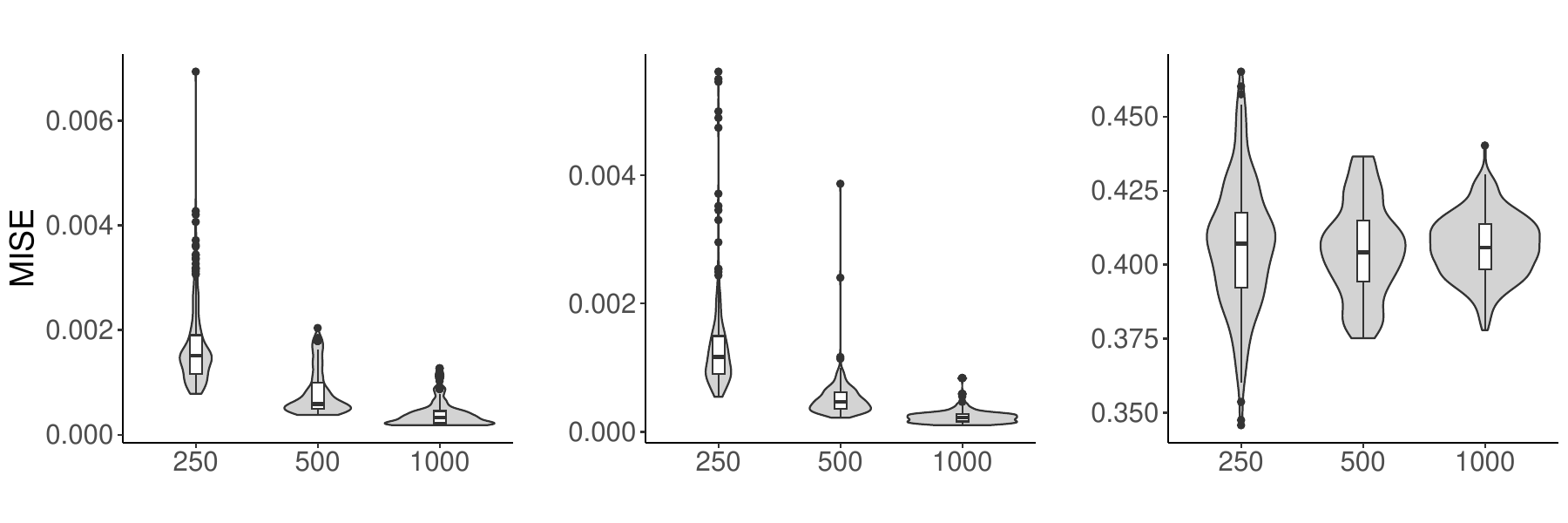} \vspace{-0.2cm}
  \includegraphics[width=\linewidth]{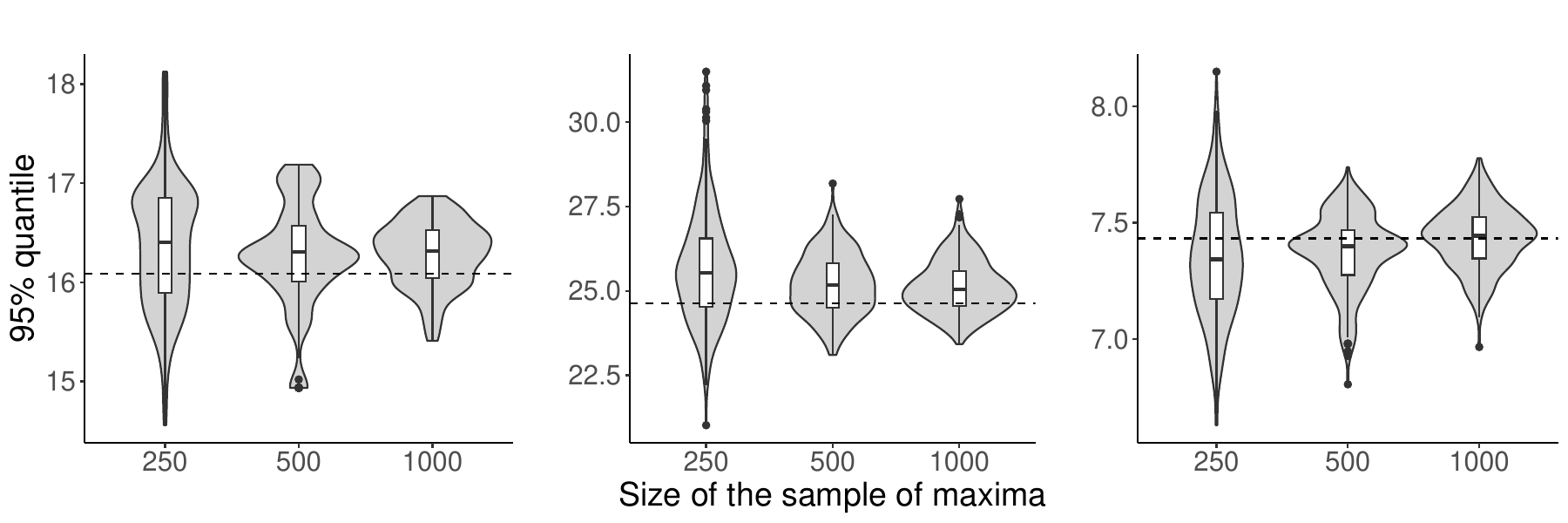}
    \caption{\footnotesize Monte Carlo simulation results, II. Top: side-by-side violin plots of MISE. Bottom: side-by-side violin plots of 95\% quantile fits from simulation experiments plotted against true 95\% quantile.}  \label{fig:MC2}  
\end{figure}

 \vc{The MISE results are reported in the top half of Fig.~\ref{fig:MC2} for each scenario, not only for the full simulated samples, but also for subsamples of smaller size. Violin plots and boxplots are shown to summarize the results across the 250 simulated datasets.}
Consistent with Fig.~\ref{fig:MC}, \vc{the plots of the MISE} confirm that the estimated densities closely match the true densities in Scenarios A and B, \vc{as the values are very small even with lower sample size. As clearly shown in the bottom panels of Fig.~\ref{fig:MC}} \vc{and already discussed,} in Scenario C the model occasionally fails to identify the correct number of components, though it still provides a reasonable estimate of the mass in the upper quantiles. \vc{Thus, it is not surprising that the MISE, which focuses on the density, is much higher in this scenario.}
Fig.~\ref{fig:MC2} also shows that, as expected, MISE decreases with increasing sample size for Scenarios A and B, confirming the good frequentist performance of the proposed method. For Scenario C, uncertainty also decreases with sample size, indicating satisfactory behavior, although a small bias persists in the limit.

The relevance of MISE in the context of extreme value analysis is, however, somewhat limited in scope. Indeed, since an extreme value analysis typically places greater emphasis on higher quantiles---and not over all values of $\mathbb{R}$---we also analyze the frequentist behavior of the proposed Bayesian approach for estimating the 95\% quantiles. \vc{We report in the bottom half of Fig.~\ref{fig:MC2} the violin plots of the posterior median 95\% quantiles, compared to the true ones, for each scenario and for each sample size}.
The obtained results again demonstrate satisfactory frequentist behavior. 
\vc{In each scenario, the model is able to accurately estimate the true quantile, even in samples with much less than the original 1000 observations. Consistent with the return plots of  Fig.~\ref{fig:MC}, in Scenarios A and B there is a very slight overestimation.
As expected, as the sample size increases, the estimates becomes closer to the true value in every scenario.}
We stress that the plot for Scenario C suggests that the estimates of the 95\% quantile are free of bias. This is because, as Fig.~\ref{fig:MC} reveals, the bias in Scenario C arises primarily around the bulk of the distribution, not in the tail.

\vc{We perform further numerical experiments with the aim of comparing the proposed model to a more classical approach, which is a single GEV model. With this aim, we simulate blocks from the two exponential distributions of Example 1, and then use the mixture model on the block maxima and a single GEV model on the maxima of the double sized blocks including both exponentials. Results can be found in the supplementary materials. We recall that the overall maximum approximately follows a single GEV distribution, and indeed return levels from the mixture model and the single GEV coincide at high return periods. However, the mixture model allows to accurately model the presence of heterogeneity, which is reflected in a gain of information at lower return periods, while still capturing well the behavior at high return periods.}

\section{Real Data Examples}\label{applications}
\subsection{Lisbon Rainfall Data}

\subsubsection*{Data Description and Motivation}
We now present an application of the proposed method to precipitation data in Lisbon, Portugal. The data consist of a series of daily precipitation measurements from December 1863 to August 2018 in Lisbon, \vc{for a total of 56\,503 records}. Data spanning from 1863 to 1940 have been digitized from the archives of the IDL (Infante D. Luiz) meteorological observatory in Lisbon. Data from 1941 to 2006 were provided by IPMA (Instituto Português do Mar e da Atmosfera), while data from April 2006 onwards were again obtained from IDL, based on digitized maps of the classical station. See \citet{valente2008early} and \citet{gallego2011trends} for details on the data.

\begin{figure}[t]
    \centering
    \includegraphics[width=\linewidth]{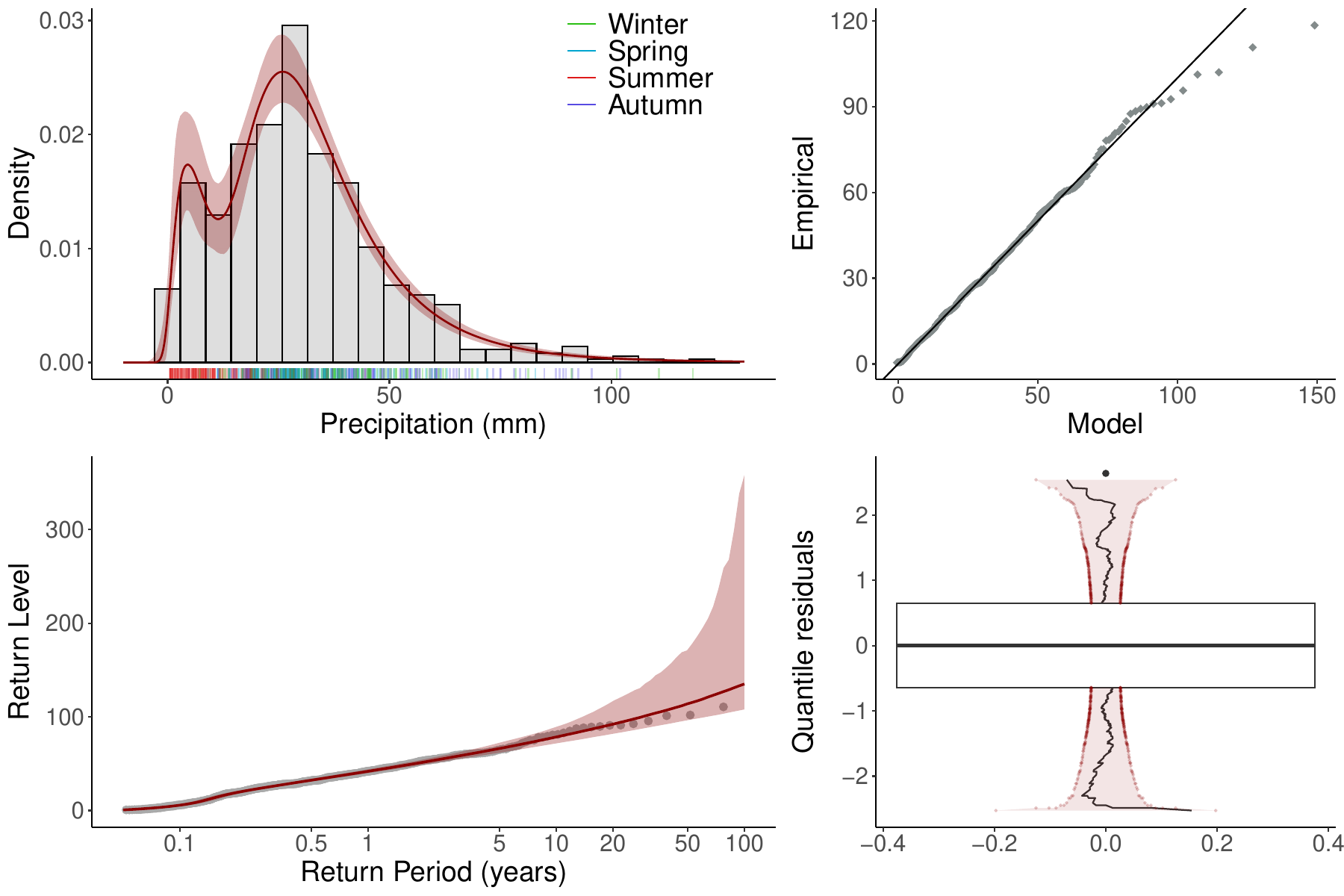}
    \caption{\footnotesize Diagnostics for model fit to seasonal precipitation maxima in Lisbon (1863–2018). Top left: median posterior density and 95\% credible intervals (shaded) overlapping the histogram of the data, with a rug plot indicating individual observations colored by season. Top right: QQ plot of posterior median quantiles. Bottom left: posterior median return level curve with 95\% pointwise credible intervals (shaded) and empirical quantiles (gray points).
    Bottom right: QQ boxplot of model residuals.}
    \label{fig:fit}
\end{figure}

Our goal is to analyze the seasonal precipitation maxima and use the proposed approach to uncover insights related to the season-driven heterogeneity in the data. We use the northern hemisphere meteorological seasons division (winter: December-February; spring: March-May;  summer: June-August; autumn: September-November). The choice of blocks of three months is guided by three considerations. Using a longer block (e.g., an entire year) would be too wide, as the model assumes that all observations within the block belong to the same group. At the same time, the block must contain enough observations for the approximation in \eqref{eq:F} to be valid. Finally, this choice is natural from an applied viewpoint as it coincides with the length of seasons. \vc{Since the data span from December 1863 to August 2018, with 154 full years (starting in December) and one year with only three recorded seasons, the sample of seasonal maxima consists of 619 observations.
The top left panel of Fig.~\ref{fig:fit} shows an histogram of the seasonal precipitation maxima, complemented by a rug plot\vc{---also informally referred to as one-dimensional scatter plots \citep[e.g.,][]{kampstra2008}---which} allows a more detailed visualization of the distribution of the data and their seasonal grouping.}

\subsubsection*{Modeling Heterogeneous Extreme Precipitation}
We fit the proposed mixture model of GEV distributions using the same MCMC setting and prior specification of the simulation study. The top left panel of Fig.~\ref{fig:fit} shows the median posterior density with corresponding pointwise credible intervals for the fitted model. Two components are recognized in the data in 99.6\% of the iterations, thus providing evidence in favor of heterogeneous extremes. \vc{Clearly, the two components are naturally associated with temporal dynamics and can be interpreted as representing different extremal regimes over time.} \vc{The rug plot that associates seasons to observations is also in line with the result from the fitted model and allows to see that the mixture components roughly correspond to observations in two different times of the year. Indeed, by comparing the two modes of the fitted density with the seasons of the corresponding observations shown in the rug plot, it is clear that} one component captures lower precipitation maxima, mainly occurring during the summer, while the other accounts for higher values, mostly associated with the remaining seasons, \vc{with the largest values occurring in winter and autumn. We would not learn as much information if we did not take into account the heterogeneity of precipitation in different seasons and instead used the classical single GEV distribution for the annual maxima, as shown in the supplementary materials.} To better understand the tail behavior, the bottom left panel of 
Fig.~\ref{fig:fit} includes the return plot with median posterior return levels and 95\% credible bands along with the empirical quantiles. 
This plot proves the ability of the fitted model to capture the heterogeneous behavior of return levels\vc{, in particular reproducing the slight curvature of the empirical quantiles around a return period of 0.2 years due to the identification of two components. Higher quantiles are also well represented, although with substantial uncertainty due to the limited number of observations in the extreme tail.}
The QQ plot in the top right panel and the QQ boxplot of model residuals in the bottom right panel confirm that the model provides an overall good fit.

\vc{In conclusion, the proposed mixture model recognizes the presence of two mixture components. Although the interest of this approach is not on classifying the observations, the two components seem to correspond to different times of the year. The flexibility of this model comes with the prize of increased uncertainty at high return periods compared to the result of a single GEV distribution fitted to the annual precipitation maxima (results available in the supplementary materials). The results are still satisfactory and, as discussed, modeling seasonal maxima allows to reveal precious insights.}

\subsection{Hong Kong Temperature Data}

\subsubsection*{Data Description and Motivation}
Another example concerns the analysis of temperature maxima in Hong Kong, and in particular seasonal maxima. Daily temperature data were recorded at the Hong Kong International Airport from January 1884 to October 2023, \vc{with a gap between January 1940 and December 1946}. A series of 532 seasonal maxima is extracted from the 48\,517 total observations \vc{covering almost 133 complete years}, using the northern hemisphere meteorological seasons division as before.

Due to the subtropical climate of Hong Kong, the area experiences distinct seasonal patterns, with hot and humid summers and a pronounced drop in temperatures during winter. This behavior creates clear heterogeneity in the seasonal maximum temperatures and motivates our analysis. The histogram of the maxima in the top-left panel of Fig.~\ref{fig:hk-fit} displays a clear bimodality, which the rug plot helps to explain: the observed winter maxima \vc{are much lower than those of the other seasons and there is a big gap between the highest winter maxima and the smallest spring maxima, creating a distinct separation.}

\subsubsection*{Modeling Heterogeneous Extreme Temperatures}

\begin{figure}[H]
    \centering
    \includegraphics[width=\linewidth]{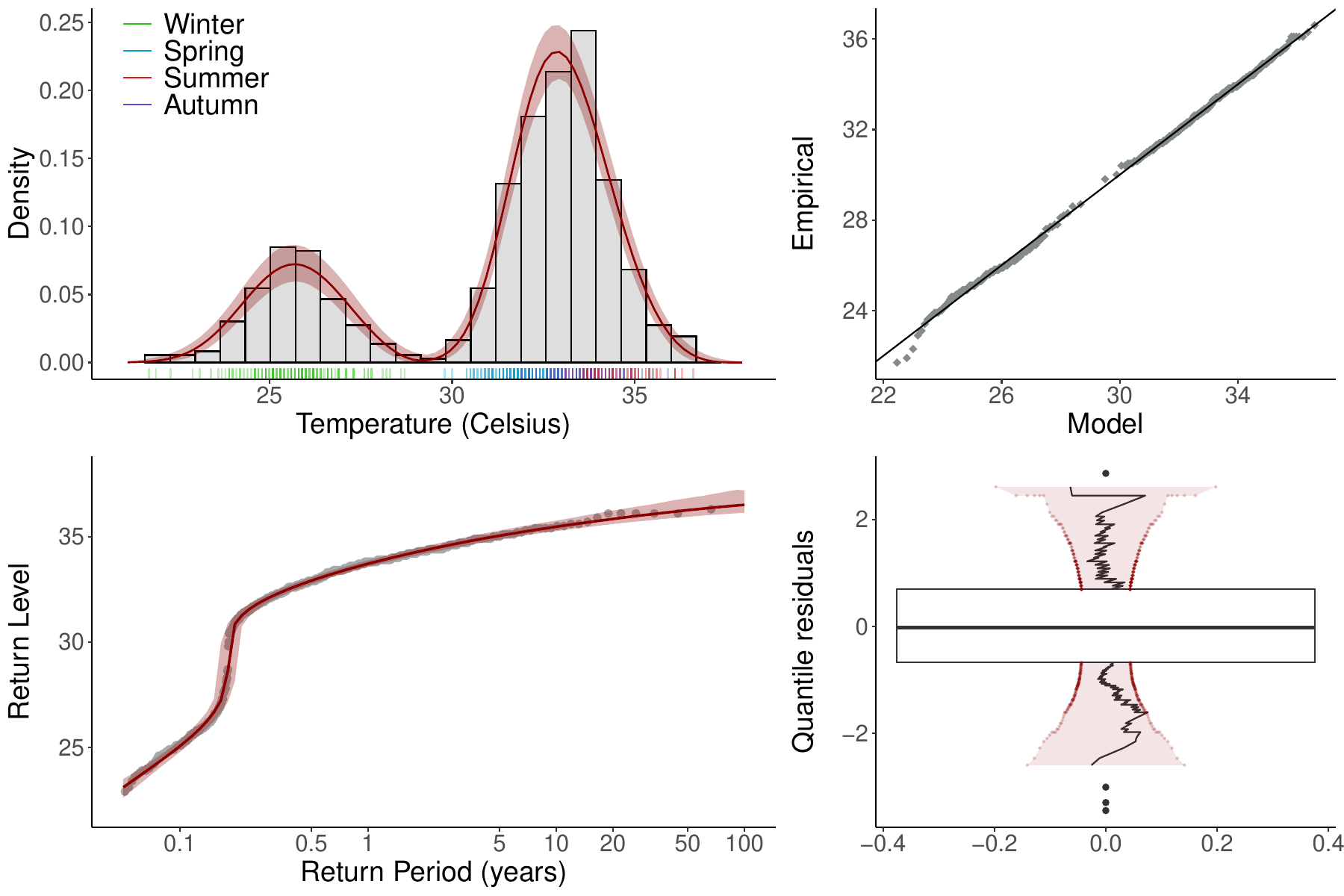}
    \caption{\footnotesize Diagnostics for model fit to seasonal temperature maxima in Hong Kong (1884–2023). Top left: median posterior density and 95\% credible intervals (shaded) overlapping the histogram of the data, with a rug plot indicating individual observations colored by season. Top right: QQ plot of posterior median quantiles. Bottom left: posterior median return level curve with credible interval (shaded) and empirical quantiles (gray points).
    Bottom right: QQ boxplot of model residuals.}
    \label{fig:hk-fit}
\end{figure}

The proposed mixture of GEV distributions is fitted using the same MCMC setup and prior specifications adopted in the simulation study. Consistent with the patterns suggested by the histogram and rug plot, in 99.9\% of the iterations \vc{of the fitted algorithm} two occupied components are identified by our fitted mixture model. The posterior median density shown in the top-left panel of Fig.~\ref{fig:hk-fit} reflects, indeed, the bimodal behavior of the histogram, confirming that the heterogeneity due to the season is captured by our model. 
The return level plot (bottom left panel of Fig.~\ref{fig:hk-fit}) shows that the fitted model effectively captures the complex tail behavior of the data. \vc{The estimated return levels well represent the empirical ones at every return period, and without a big increase in uncertainty at the highest quantiles}. Additional diagnostic plots, displayed in the right panels, further confirm the adequacy of the model. \vc{The QQ plot establishes that the estimated quantiles well represent the empirical ones, and the QQ boxplot shows that model residuals are in line with the Normal distribution}.

\vspace{-0.2cm}The data contain only 110 unique values, likely due to recording precision. This does not affect the fit. Nonetheless, motivated by the large number of ties, we devised an interval-censored approach that can be used to formally adjusting the proposed methodology for settings with ties or rounding errors; see Appendix~B for details. The fitted density stemming from such interval-censored approach is, however, visually indistinguishable from that in Fig.~\ref{fig:hk-fit} and is therefore reported only in the supplementary materials.

\vc{A competing and more classical approach when dealing with temperature data is simply to take the annual maxima and fit a single GEV distribution, for instance using a Metropolis--Hastings algorithm; the results of this analysis are shown in the supplementary materials. The typical approach entails losing all the information about the seasonal pattern of temperatures, as well as losing precision in estimating high quantiles. By considering seasonal maxima and fitting a mixture model, we are able to capture the bimodal nature of the seasonal block maxima---clearly visible in the rug plot in Fig.~\ref{fig:hk-fit}---and to obtain reliable estimates of the return levels. Finally, it is worth recalling that, as emphasized in Section~\ref{sec:2.1}, the choice of block size is not arbitrary: it must be sufficiently large for the asymptotic justification of extreme value models to hold and, in practice, is typically chosen to have a meaningful interpretation for the application (e.g., seasonal).}

\section{Closing Remarks}
\label{sec:Final}
This work pioneers the statistical modeling of heterogeneous extremes by extending traditional block maxima methods to settings with cross-block heterogeneity. Our extreme value mixture model mitigates misspecification risk by allowing each block maximum to arise from a distinct GEV component; our approach includes the GEV as a special case while accommodating deviations when data exhibit cross-block heterogeneity.

Grounded in the extremal types theorem and its formulation in \eqref{eq:F}, the proposed mixture model introduces flexibility by adapting to varying degrees of heterogeneity and capturing complex distributional shapes when required. The infinite sum case can be formally addressed, as shown in Appendix~A, by combining the law of total probability with a well-known dominating convergence-type theorem.

We thus question the common practice of defaulting to the GEV distribution solely because it arises as a theoretical limit. As shown, alternative block-maxima-based models can be constructed that remain consistent with the extremal types theorem while offering greater robustness to misspecification and adaptability to diverse data patterns.

One may wonder about the computational overhead of fitting the proposed approach. Given state-of-the-art expertise in fitting infinite mixtures \citep[e.g.,][]{ishwaran2002, kalli2011}, such models are relatively straightforward to fit and are standard practice in the nonparametric Bayesian literature. While the GEV kernel presents some challenges compared to the normal kernel \citep[e.g., the lack of conjugacy prevents the use of a blocked Gibbs sampler, as in][]{ishwaran2002}, these do not pose major obstacles, as detailed in the supplementary materials.

Several extensions could be explored in future work. Firstly, a natural direction is to  incorporate covariates into the mixture model using the dependent Dirichlet process (DDP) or its extensions, as reviewed by \cite{quintana2022dependent}. This approach would lead to the following extension:
\begin{equation}\label{extensionddp}
F(z \mid \mathbf{x}) = \sum_{k=1}^\infty \pi_k G(z\mid \mu_k(\mathbf{x}), \sigma_k(\mathbf{x}), \xi_k(\mathbf{x})),
\end{equation}
where $\mu_k(\mathbf{x})$, $\sigma_k(\mathbf{x})$, and $\xi_k(\mathbf{x})$ are generalized linear models; see, for example, \cite{inaciodecarvalho2017} for a related specification based on the Gaussian kernel instead of the GEV.  Eq.~\eqref{extensionddp} can be interpreted as a mixture-of-experts extension of a mainstream regression approach for extremes based on setting $F(z \mid \mathbf{x}) = G(z \mid \mu(\mathbf{x}), \sigma(\mathbf{x}), \xi(\mathbf{x}))$  \citep[see][Chapter~6]{coles2001}. Secondly, it is natural to consider the possibility of allowing for heterogeneity at the level of exceedances, which, using similar arguments as in Section~\ref{methods}, suggests that a mixture of Generalized Pareto models could be asymptotically justified for heterogeneous exceedances.  While mixture models based on the Generalized Pareto distribution have been considered elsewhere \citep[e.g.,][]{cabras2011}, they neither account for heterogeneous exceedances nor provide a formal EVT-based motivation in the sense developed here. Finally, the proposed framework naturally raises questions about extensions to multivariate and spatial settings. While the univariate case developed here already presents significant challenges, extending the framework to more complex structures would further advance our understanding of heterogeneous extremes; such developments could also relate with and extend a recent spatial extremes mixture model by \cite{richards2023}. \vc{It is also worth mentioning that the components of our model could in some cases be associated with temporal effects (e.g., a winter effect in Fig.~\eqref{fig:hk-fit}) and thus be interpreted as representing distinct extremal regimes over time. This perspective naturally raises the possibility of linking components to structural change points; exploring such connections constitutes an interesting direction for future research, particularly in view of the relatively limited literature in this area \citep{dierckx2010,de2020,lee2021}. Finally, we emphasize the scope for formal large–sample characterizations of the posterior in our mixture GEV model with heterogeneous block maxima. Even in the single-component case,  this question is rather challenging \citep{padoan2024, carl2025}. A main difficulty is that the GEV distribution is only visible in the limit, so in finite samples each block maximum is at best approximately GEV–distributed; the same applies to our heterogeneous block maxima approximation. We leave these open questions for future research.} 

\appendix \footnotesize 
\section*{Appendix: Technical Details \& Proofs}\label{details}
\renewcommand{\theequation}{A.\arabic{equation}}
\setcounter{equation}{0}
\subsection*{Appendix A.~Details on Extending \eqref{eq:F} for the Case $K \to \infty$}\label{details}
First, we recall the so-called Tannery's lemma \citep{tannery1910introduction}.
\begin{lemma}\label{tannery}
  If $|C_{k, n}| \leq D_k$, for all $k \in \mathbb{N}$ and $n \in \mathbb{N}$, and if $\sum_{k = 1}^{\infty}D_k$ converges, then
  \begin{equation*}
    \lim_{n \to \infty} \sum_{k = 1}^\infty C_{k, n} =  \sum_{k = 1}^\infty \lim_{n \to \infty}  C_{k, n},
  \end{equation*}
  provided that the limits on the right-hand side exist. 
\end{lemma}
\noindent Next, we show that \eqref{eq:F} can be extended for the case $K \to \infty$ by relying on Lemma~\ref{tannery}. To show this, we first rewrite the definitions of the group-dependent normalizing sequences of random variables in \eqref{gdepcons} for this case as 
\begin{equation*}
  A_n = \sum_{k = 1}^{\infty} I(C = k)\,a_{k, n}, \qquad  B_n = \sum_{k = 1}^{\infty} I(C = k)\,b_{k, n}.
\end{equation*}
Let $C_{k, n}(z) = \pi_{k} \text{P}\{(M_{k, n} - b_{k, n}) / a_{k, n} \leq z\}$ and note that $C_{k, n}(z) \leq \pi_k \equiv D_k$, for all $z \in \mathbb{R}$. Clearly, $\sum_{k = 1}^\infty D_k$ converges (indeed, $\sum_{k = 1}^\infty D_k = 1$). Next, in a similar fashion as in \eqref{eq:1F}, it follows from the law of total probability that 
\begin{equation}\label{eq:2Fa}
  \text{P} \left ( \frac{M_n-B_n}{A_n}  \le z \right) = \sum_{k=1}^\infty \text{P}(C=k) \, \text{P} \left ( \frac{M_n-B_n}{A_n}  \le z  \mid C=k \right ) 
 = \sum_{k=1}^\infty \pi_k \, \text{P} \left ( \frac{M_{k,n}-b_{k,n}}{a_{k,n}}  \le z \right). 
\end{equation}
Hence, it follows from \eqref{eq:2Fa}, by Lemma~\ref{tannery}, and the extremal types theorem that
\begin{equation*}
  \begin{split}
    \lim_{n \to \infty} \sum_{k = 1}^{\infty} \pi_k \text{P}\bigg(\frac{M_{k, n} - b_{k, n}}{a_{k, n}} \leq z\bigg) 
    &=  \lim_{n \to \infty} \sum_{k = 1}^{\infty} C_{k, n}(z)  \\ 
    & =  \sum_{k = 1}^{\infty} \lim_{n \to \infty} C_{k, n}(z) \\ 
    &=  \sum_{k=1}^{\infty} \lim_{n \to \infty} \pi_k \,\text{P} \left ( \frac{M_{k,n}-b_{k,n}}{a_{k,n}}  \le z \right) \\ 
    & = \sum_{k=1}^\infty \pi_k \,G_{\xi_k}(z),     
  \end{split}
\end{equation*}
pointwisely, for any $z \in \mathbb{R}. \strut \hfill \square$

\renewcommand{\theequation}{B.\arabic{equation}}
\setcounter{equation}{0}
\subsection*{Appendix B.~Details on Ties and Rounding Issues}\label{ties}
In some cases, repeated values may arise due to rounding, but the proposed methodology can be readily adapted for that context. Suppose that, rather than knowing the exact data point $z_i$, we only observe that $z_i$ falls within the interval $(z_i^L, z_i^R]$. Thus, the likelihood contribution for an interval-censored observation from a GEV distribution is determined by the probability of being within the given interval:
\begin{equation*}
    \text{P}(z_i^L < z_i \le z_i^R \mid \thetab_k) = G(z_i^R \mid \boldsymbol{\theta}_k) - G(z_i^L\mid \boldsymbol{\theta}_k),
\end{equation*}
where  $\thetab_k = (\mu_k, \sigma_k, \xi_k)^{\T}$. It follows that the contribution to the complete data likelihood stemming from censored GEV data belonging to group $k$ is, \begin{equation}\label{eq:lik-cens}
    g_{\text{cens}}(\bm{z} \mid \thetab_k) = \prod_{\{c_i = k\}} [G(z_i^R\mid \boldsymbol{\theta}_k) - G(z_i^L\mid \boldsymbol{\theta}_k)],
  \end{equation}
  where $c_1, \dots, c_m$ are allocation indicators. Treating rounded data as a specific case of interval censored data implies that $z_i^L= z_i^D-\delta$ and $z_i^R= z_i^D + \delta$, with $\delta>0$, where $z_i^D$ is the recorded rounded data point. In particular, $\delta$ depends on the granularity of the measurement scale and it is determined by the precision to which the data are recorded. Indeed, an observation represents all true values within an interval, and $\delta$ determines how far on either side of a recorded value the true value could be. For example, if data are rounded to the first decimal digit, an observation like 2.3 corresponds to the interval (2.25, 2.35], and $\delta=0.05$. 

  In the supplementary materials, we provide detailed comments on posterior sampling, including guidance on handling this interval-censored setting. In broad strokes, the important things to keep in mind are that the interval-censored likelihood that follows from \eqref{eq:lik-cens} should be used, and the allocation indicators are sampled according to the corresponding interval-censored structure. In particular, the allocation indicators are sampled based on
\begin{equation}\label{aisample}
    \text{P}\{C_i = k \mid z_i \in (z_i-\delta, z_i + \delta]\} = \frac{\pi_k \{G(z_i+\delta \mid \thetab_k) - G(z_i-\delta\mid \thetab_k)\}}{\sum_{j=1}^K \pi_j \{G(z_i+\delta \mid \thetab_j) - G(z_i-\delta\mid \thetab_j)\}},
  \end{equation}
  for $i = 1, \dots, m$.

\section*{\centering Acknowledgment}
\begin{centering}\scriptsize 
We thank the Editors and two anonymous reviewers for their insightful comments on an earlier version of this paper. 
We also thank Daniela Castro Camilo and Raffaele Argiento for valuable feedback on a previous draft, and the participants of ICSDS 2024 and RISE 2025 for helpful comments. VC was funded by the FIRE-RES project within the H2020 program of the European Union (EU), and later by the project CLIMATHS of the PEPR Maths-Vives under the France 2030 investment plan.
MdC was supported by RSE, Leverhulme Trust, UnaEuropa, and CIDMA (FCT UIDB-UIDP/04106/2020). IP and IAV acknowledge the support of the RISE project (NextGenerationEU---National Recovery and Resilience Plan, 
PRIN–CUP N.H53D23002010006). Finally, this work was supported by DoE 2023-2027 (MUR, AIS.DIP.ECCELLENZA2023\_27.FF project). 
This work reflects only the authors’ views and opinions; neither the EU nor the European Commission can be considered responsible for them.  
\end{centering} \vs \vs 

\noindent \textbf{Declarations} \\
\noindent \textbf{Conflict of Interest}: The authors declare that they have no conflicts of interest relevant to the content of this article. \vs 

\noindent \textbf{Data Availability Statement}: 
The Lisbon and Hong Kong datasets used in Section~\ref{applications} were retrieved from the \textsf{R} package \texttt{DATAstudio} \citep{datastudio}. 

\bibliographystyle{asa2.bst} 
\bibliography{bibliography}       


\end{document}

%% file: colors.tex
\usepackage{color}



%% file: environments.tex
\newtheorem{example}{Example}

\newtheorem{lemma}{Lemma}


%% file: tools.tex
\newcommand{\T}{\ensuremath{\mathrm{\scriptscriptstyle T}}}
\usepackage{bm}

\newcommand{\thetab}{\ensuremath{\bm\theta}}

\usepackage{amsmath}

\DeclareMathOperator{\iid}{\overset{iid}{\sim}}

\usepackage{mathrsfs}

\newcommand{\vs}{\vspace{0.2cm}}


%% file: short_bib.tex
 \let\oldthebibliography=\thebibliography
 \let\oldendthebibliography=\endthebibliography
 \renewenvironment{thebibliography}[1]{%
   \footnotesize
   \oldthebibliography{#1}%
   \setlength{\parskip}{0mm}%
   \setlength{\itemsep}{0mm}%
 }{\oldendthebibliography}
